\documentclass [12pt]{article}
\usepackage{hyperref}

\usepackage{graphicx}

\def\beq{\begin{equation}}
\def\eeq{\end{equation}}
\def\bea{\begin{eqnarray}}
\def\eea{\end{eqnarray}}

\begin{document}

\title { $^{10}$Li spectrum from $^{11}$Li fragmentation.}

\author { G. Blanchon$^{(a)}$, A. Bonaccorso$^{(a)}$, D. M.  
Brink$^{(b)}$,
 and N. Vinh Mau$^{(c)}$\\
\small $^{(a)}$ Istituto Nazionale di Fisica Nucleare, Sez. di
Pisa, \\
\small and Dipartimento di Fisica, Universit\`a di Pisa,\\
\small Largo Pontecorvo 3, 56127 Pisa, Italy.\\
\small $^{(b)}$ Department of Theoretical Physics, 1 Keble Road, Oxford  
OX1 3NP, U. K.\\
\small $^{(c)}$ Institut de Physique Nucl\'eaire, IN2P3-CNRS, F-91406,  Orsay Cedex, France. }

\maketitle

\begin{abstract}

A recently developed  time dependent model for the excitation of a  
nucleon from a bound state  to a continuum resonant state in  the system n+core  is applied to the study  of the population of the low  
energy continuum of  the unbound  $^{10}$Li system obtained from  $^{11}$Li fragmentation. 
Comparison of the model results to new data from the GSI laboratory suggests  that  the reaction mechanism  is dominated by final state effects rather than by the sudden process,  but for the population of the l=0 virtual state, in which case the two mechanisms give almost identical results.  There is also, for the first time, a clear evidence for the population of a d$_{5/2}$ resonance in $^{10}$Li. \end{abstract}

\section{Introduction}

In this paper we apply the {\it projectile fragmentation} model developped in Ref. \cite{1} to the  the elastic breakup (diffraction reaction) of the two-neutron halo nucleus $^{11}$Li on a $^{12}$C target, as recently measured at GSI and presented in Ref. \cite{2}. The observable studied is the one-neutron-core relative energy spectrum.  These type of data are necessary to enlighten the effect of the neutron final state interaction with the core of origin. 
 
Light unbound nuclei have  attracted much attention  
\cite{bhe}-\cite{Labi99} in connection with exotic halo nuclei. Besides, a precise
understanding of unbound nuclei is essential to determine the position  
of the driplines in the nuclear mass chart. In two-neutron halo nuclei  
such as $^{6}$He, $^{11}$Li, $^{14}$Be,  the two neutron pair  
is bound,
although weakly, due to the neutron-neutron pairing force, while each  
single extra neutron is unbound in the field of the core.
In a three-body model these nuclei are described as a core plus two  
neutrons. The properties of the  core plus one neutron system are
essential to structure models which rely on the knowledge of single particle quantum numbers such as angular  momentum, parity, energies, as well as the corresponding neutron-core
effective potential. With those ingredients the spectroscopic strength function of neutron  
resonances in the field of the core can be obtained.

We will use the projectile fragmentation formalism, an inelastic-like
excitation to the neutron-core continuum, to the study of the effect of final state interaction of the neutron with the  
projectile core.  The model  is a  theory which has already been shown to be relevant  to the interpretation of neutron-core coincidence  
measurements in nuclear elastic breakup reactions with projectiles of $^{14}$B and $^{14}$Be \cite{1,2}. In the case of two nucleon breakup we describe  only the step in which a neutron is knocked out from the projectile by the neutron-target interaction to first order and then re-interacts in the final state with the core. The case in which a resonance is populated by a sudden process while the other neutron is stripped has been already discussed in Ref. \cite{bhe} and it has been shown  \cite{1}  that there is   a simple link between the two methods of Ref. \cite{1} and Ref.\cite{bhe}.  We assume that the neutron which is not detected has been stripped while the other suffers an elastic scattering on the target. The influence of the second nucleon is taken into account only by a modification
of the neutron-core interaction in the final state.  Section 2 contains a brief reminder of the formalism from Ref. \cite{1}. Section 3 describes the results of our numerical calculations
for $^{11}$Li  which, being already well understood \cite{2}-\cite{menic},  is used here  as  a test case.  Details on our assumptions for the potentials needed in the calculations are also presented.   Finally our conclusions are contained in Sec.   4.

 \section {Formalism for inelastic excitation to the continuum. }

\begin{figure}[h!b]
         \scalebox{0.3}{
              \includegraphics{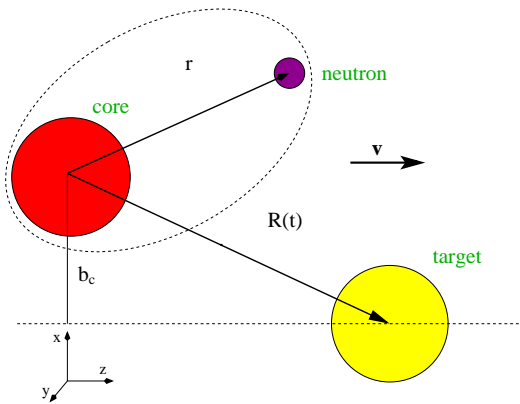}}  
              \caption{Coordinate system used in the calculations}                  
  \label{refsys}     \end{figure}
Following  Refs. \cite{1,bb1,adl,bw}, we will describe the inelastic-like excitations of one neutron, from a  bound state $\psi_i$ to a final state  
$\psi_f$  in the continuum, by the
time dependent
perturbation amplitude :

\begin{equation}A_{fi}={1\over i\hbar}\int_{-\infty}^{\infty}dt \langle \psi_{f} ({\bf r},t)|V_2({\bf { r-R}}(t))|\psi_{i}({\bf r},t)\rangle.\label{1}
\end{equation}
 $V_2$ is the interaction responsible for the neutron transition (cf. Eq.   (2.15) of \cite{bb1}).  The potential $V_2({\bf { r-R}}(t))$
 moves past on a constant velocity path with velocity $v$ in the z-direction with an impact parameter $b_c$ in
the x-direction in the plane $y = 0$. This assumption  makes our semiclassical model valid at beam energies well above the Coulomb barrier. This is in fact the regime in which projectile fragmentation experiments are usually performed (cf. Sec.   3).
The coordinate system used in the calculations is shown in Fig. \ref{refsys} and it corresponds to the no-recoil approximation for the core. 
Let  $\psi_{i} ({\bf r},t)=\phi_{i}({\bf r})e^{-{i\over \hbar}\varepsilon_it}$ be the single particle initial state wave function. Its  radial part $ \phi_{i}({\bf r})$ is calculated in a potential $V_{WS}( r)$ (cf. Sec.   3.1) which is  
fixed in space.  
  For initial and final states of different  angular momentum  our wave functions are trivially orthogonal due to the orthonormality of their angular parts. For transitions conserving the angular momentum of the single particle states the  orthogonalization correction has been estimated in Ref. \cite{1} and found negligible.  
  Now change variables and put $z - vt = z^{\prime}$ or $t = (z -  
z^{\prime})/v$, define 
\begin{equation}
q={{\varepsilon_f-\varepsilon_i}\over {\hbar v}},\label{q}
\end{equation}
then
choosing  
 $V_2(r) $ to be a delta-function potential
$V_2(r) = v_2\delta(x)\delta(y)\delta(z)$, with $v_2\equiv$ [MeV fm$^3$],   the integrals over x and y can be calculated giving
\begin{equation}
A_{fi}={v_2\over i\hbar v}
\int_{-\infty}^{\infty}dz ~ \phi^*_{f} (b_c,0,z)\phi_{i}(b_c,0,z){e^  
{iqz}}.\label{1bis}
\end{equation}
The value of the strength $v_2$ used in the calculation is  discussed in Sec.   3.

The following
forms for the wave functions will be used. For the initial bound state

\begin{equation}
\phi_{i}({\bf r})=-C_ii^{l_i}\gamma h^{(1)}_{l_i}(i\gamma  
r)Y_{{l_i},{m_i}}(\theta,\phi).\label{f}
\end{equation}
Due of the strong core absorption implied in the following by Eq.  (\ref{cross}), we use in
this paper the asymptotic form of the initial state wave function, given in terms of the Hankel functions $h^{(1)}_{l}$.
However the exact wave function, numerical solution of the bound
state Schr\"odinger equation can be used without introducing further  
complexity in the calculations. For the final continuum state

\begin{equation}
\phi_{f}({\bf r})=C_f k{i\over 2}(h^{(-)}_{l_f}(kr)-S_{l_f}  
h^{(+)}_{l_f}(kr))Y_{{l_f},{m_f}}(\theta,\phi),\label{p2}
\end{equation}
 $C_f=\sqrt{2/L}$ is the
normalization constant for the  final state. L  
is a large box radius used to normalize the continuum wave function  
(cf. Eq. (2.5) of Ref. \cite{bb}). $S_{l_f}=e^ {2i\delta_{l_f}}$ is the neutron-target S-matrix in the $l_f$ partial wave.

The probability to excite a final continuum state of energy  
$\varepsilon_f$ is an average over the initial state and a sum over the final states. Thus introducing the density of final states and the density of final states, according to Ref. \cite{bb}
$
\rho (\varepsilon_f) d\varepsilon_f={L m/  \pi \hbar^2 k}  
d\varepsilon_f,$
the probability spectrum reads
 
\begin{eqnarray}
{dP_{in}\over d\varepsilon_f}={2\over \pi}{v_2^2\over \hbar^2  
v^2}{C_i^2 }{m\over\hbar^2k}{1\over 2l_i+1}\Sigma_{m_i,m_f}
 |1-\bar S_{m_i,m_f}|^2 |I_{m_i,m_f}|^2, \label{8}
\end{eqnarray}
where now  $
\bar S=S e^{2i\nu}=e^{2i(\delta+\nu)}
$ is an off-the-energy shell S-matrix and $\nu$ is the phase of the integral:
\begin {equation} I_{m_i,m_f}=\int_{-\infty}^{\infty} dze^{iqz}i^{l_i} \gamma h^{(1)}_{l_i}(i\gamma  
r)Y_{{l_i},{m_i}}(\theta,0)  k{i\over 2}h^{(-)}_{l_f}(kr)Y_{{l_f},{m_f}}(\theta,0).\label{8bis} \end{equation} 
A detailed derivation of the above equations can be found in Sec.2 of  Ref. \cite{1}. For simplicity the equations in this paper are given without  spin variables in the initial and final states. The generalization including spin is given in Appendix B of Ref. \cite{1}. 

The cross section
differential in $\varepsilon_f$ 
is given as

\begin{equation}
{d\sigma_{-1n}\over {d{{\varepsilon_f}} }}=C^2S
\int d{\bf b_c} {d P_{in}(b_c)\over
d{{\varepsilon_f}}}
P_{ct}(b_c), \label{cross}
\end{equation}
(see Eq.   (2.3) of \cite{bb4}) and C$^2$S is the spectroscopic factor  
for the initial state.

The core survival probability $P_{ct}(b_c)=|S_{ct}|^2$ \cite{bb4} in  
Eq.  (\ref{cross}) takes into account the peripheral nature of
the reaction and naturally excludes the possibility of large overlaps  
between projectile and target. $P_{ct}$ is defined in terms
of a S-matrix function of the core-target distance of closest approach  
$b_c$. 
A simple parameterisation is
$P_{ct}(b_c)=e^{(-\ln 2 exp[(R_s-b_c)/a])}$ \cite{bb4}, where the  
strong absorption radius R$_s\approx1.4(A_p^{1/3}+A_t^{1/3})$ fm is  
defined
as the distance of closest approach for a trajectory that is 50\%  
absorbed from the elastic channel and a=0.6 fm  is a diffuseness
parameter. 

\section{Applications }

\subsection{One neutron average potential }

We apply  the fragmentation model to the study of the relative energy spectrum  n+$^{9}$Li obtained by the authors of Ref. \cite{2} in the breakup reaction of $^{11}$Li on $^{12}$C at 264 A.MeV. The structure of $^{11}$Li is already well known from a number of experiments and theoretical papers \cite{2}-\cite{menic}: the two neutron separation energy  is  0.3$\pm$0.27 MeV \cite{nucbase}; the wave function is combination of a 2s state with a spectroscopic factor  0.31, a p$_{1/2}$  with spectroscopic factor 0.45 \cite{ers} and there is also a small d$_{5/2}$ component. The main d$_{5/2}$ strength is in the continuum centered around 1.55 MeV \cite{2}. The link between reaction theory and structure model is made by the  
neutron-core potential determining the S-matrix in Eq.  (\ref{p2}).
Then if the theory fits the position and shape of the continuum  
n-nucleus energy distribution, obtained for example by a coincidence
measurement between the neutron and the core,  the parameters of a  
model potential  can be deduced.  

The calculations in the present paper are made with different  
potentials for the initial
and final state.   
 The initial wave function for the s-state is calculated in a simple Woods-Saxon potential with R= r$_0A^{1/3}$ and  strength fitted to the separation energy 0.3 MeV\cite{nucbase} and whose parameters are given in Table \ref{o}.
\begin{equation}V_{WS}(r)={V_0\over{1+e^{(r-R)/ a} }}-\left  
({\hbar\over m_{\pi}c}\right )^2{V_{so}\over ar}{e^{(r-R)/
a}\over {(1+e^{(r-R)/ a})^2}}{\mathbf {l \cdot \sigma}}\label{n}
\end{equation}
 
   \begin{table}[h]
\caption {\footnotesize Asymptotic normalization constants C$_i$(fm$^{- {1/2}}$) for the initial state components wave functions of the bound neutron. Spectroscopic factors from Ref. \cite{ers}.  Potential parameters are:  
V$_0$  fitted to give the  separation energy 0.3 MeV  \cite{nucbase}. The  
other potential parameters are r$_0$=1.27 fm,
a=0.75 fm,  V$_{so}$=5.25 MeV.   }
\begin{center}\footnotesize \begin{tabular}{c|cc}
\hline
\hline
$\varepsilon_i $&~~~-0.3 MeV&
\\ \hline 

$l_i,~~ j_i$&  C$_i$(fm$^{- {1/2}}$)&C$^2$S \\ \hline
 0 $~~{1/2}$   &0.76&0.31 \\
 1 $~~{1/ 2}$   &0.24& 0.45 \\
\hline\hline
\end{tabular}\end{center}\label{o}
\end{table}

\begin{table}[hb]
\caption {\footnotesize Woods-Saxon and spin-orbit potential
parameters for the continuum final states.}
\begin{center}
\footnotesize
\begin{tabular}{ccccc}
\hline
\hline
  V$_0$  &  r$_0$  &  a   & V$_{so}$ & a$_{so}$ \\
(MeV) &  (fm) & (fm) & (MeV)  &  (fm) \\
\hline
-39.8   &   1.27  & 0.75   &   7.07 & 0.75 \\
\hline
\hline
\end{tabular}
\end{center}\label{p}
\end{table}

To describe the valence neutron in the $^{10}$Li continuum we
assume that the single neutron hamiltonian with respect to $^{9}$Li  
has the form
\begin{equation}
h=t+ U+iW
\label{op}\end{equation}
where $t$ is the kinetic energy and
\begin{equation}
U (r) =V_{WS}+\delta V \label{pot}\label{pot0}
\end{equation}
is the real part of the neutron-core interaction. In this paper the
imaginary part is taken equal to zero. V$_{WS}$ is again a Woods-Saxon  
potential plus spin-orbit whose parameters are given in Table \ref{p},
and $\delta$V is a correction \cite{Vinh96}:
\begin{equation}
\delta V(r)=16\alpha {e^{{{2(r-R)}/ a}}/({1+e^{{{(r-R)}/ a}}})^4  
}\label{pot1}
\end{equation}
which originates from particle-vibration  
couplings. They are important for low energy states but can be
neglected at higher energies.  The above form  is suggested by a calculation of such couplings using
 Bohr and Mottelson collective model of the transition amplitudes between zero and one phonon
states. Therefore our structure model is not a simple {\it single-particle in a potential} model but contains in it the full complexity of single-particle vs. collective couplings.  A more realistic treatment would require the description of both bound and unbound states in a three-body model such as in Refs. \cite{rod} and \cite{IJT}. 

The continuum energies can be adjusted by varying the parameter  
$\alpha$ in the potential.
By changing the strength $\alpha$  
of the $\delta$V potential in Eq.  (\ref{pot1}) we could make also  the  2s-state just bound  
near threshold and see
 what would be  
  the effect on the continuum spectrum (cf.\cite{1,bbv}). 
  As possible final states we have considered only the s, p and d partial waves calculated in the potential of Table \ref{p} plus Eq.  (\ref{pot1}), according to
Ref. \cite{Vinh96}, with
different values of the strength $\alpha$. 
Table \ref{q1}  gives  the energies and widths of the 1p$_{{\footnotesize {1/2}}}$ and  
1d$_{5/2}$ states and the corresponding
 values of  $\alpha$. The widths are obtained from  
the phase shift  variation near  resonance energy, according to
$d\delta_{j}/d\varepsilon_f|_{\varepsilon_{res}}=2/\Gamma_j$, once that  
the
resonance  energy is fixed \cite{joa}. Notice that the values of $\alpha$
for the s and d states are very similar, therefore these two states are basically obtained in the same potential.
\begin{table}[ht]
\caption{\footnotesize Scattering length of the 2s continuum state, energies and widths of the p- and d-resonances  
in $^{10}$Li  and corresponding strength parameters for the $\delta$V  
potential.}\vskip .1in
\begin{center}
\footnotesize
\begin{tabular}{lccccc}
\hline
\hline\
&&$\varepsilon_{res}$ &$\Gamma_j$& a$_s$&$\alpha$\\
&&MeV &MeV&fm$^{-1}$ & MeV\\
\hline
&2s$_ {1/2}$ & & &-17.2&-10.0\\
&1p$_ {1/2}$ & 0.63&0.35 &&3.3\\
&1d$_{5/2}$ & 1.55 &0.18&&-9.8\\
\hline
\hline
\end{tabular}
\label{q1}\end{center}
\end{table}

 The delta-interaction strength used in Eq.  (\ref{8}), is $v_2$=-8625 MeV fm$^3$. It has been obtained by imposing that this interaction gives the same volume integral as a n-$^{12}$C Woods-Saxon potential of strength -50.5MeV, radius 2.9 fm and diffuseness 0.75 fm.

\subsection {Results }

\begin{figure}[h!t]
\center
 \vskip 20pt
 \scalebox{0.4}{
               \includegraphics{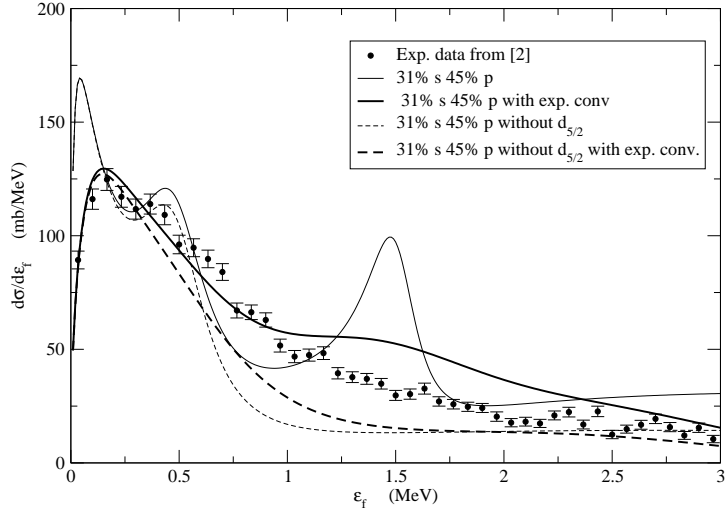}}
\caption{ \footnotesize n-$^{9}$Li relative energy  spectrum,
 for the reaction  $^{11}$Li+$^{12}$C $\to$ n+$^{9}$Li+X at 264 A.MeV. Only the contributions from an s and p initial state with  experimental 
spectroscopic factors \cite{ers} C$^2$S= 0.31  and 0.45 respectively are  included. The thin solid curve is the total calculated result. The thick solid curve curve is after convolution with the experimental resolution function. The thin dashed curve is the calculation without the d-resonance while the thick dashed curve is the same  calculation after convolution. The symbols with error bars are the experimental points from \cite{2}. Calculations are normalised to the data.}
\label{11Be}
\end{figure}

\begin{figure}[h]
 \vskip 20pt
\center  \scalebox{0.4}
{\includegraphics[ angle=-0]{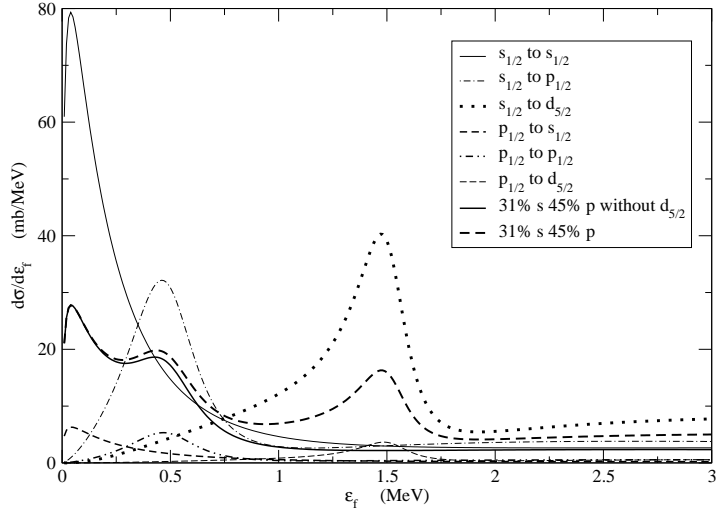}}
\caption {\footnotesize The individual transitions, bound to unbound from the s and p initial state components, taken with unit spectroscopic factors, to the s,p and d unbound state as indicated. The thick dashed line is the sum of all transitions, including  the spectroscopic factors for the initial state components while the thick solid line is does not include transitions to the d-resonance. }
\label{fig3b}
\end{figure}

\begin{figure}[h]
 \vskip 20pt
\center  \scalebox{0.4}
{\includegraphics[ angle=-0]{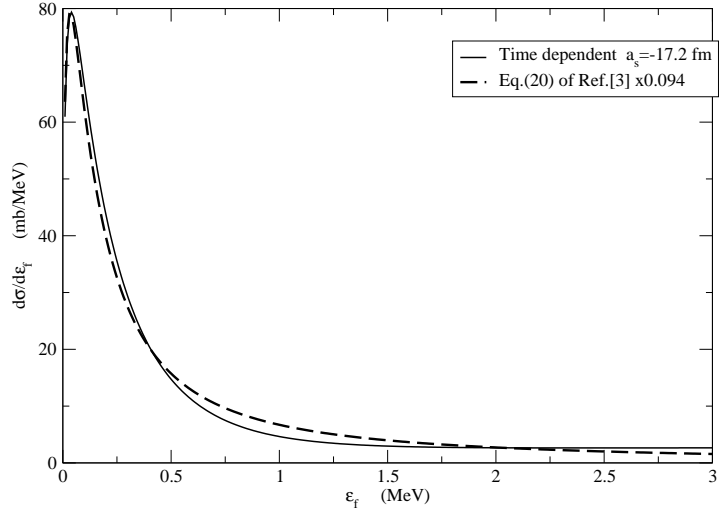}}
\caption {\footnotesize Comparison for the s to s transition of our calculation (solid curve) with that  according to the sudden formula Eq.   (20) of \cite{bhe}. Both calculations use the {\it exact} phase shifts. }
\label{fig4}
\end{figure}

Results obtained with the model outlined in Secs. 2  and 3.1  will now be  
discussed. We consider the knockout of a single neutron from a bound state in a potential, similarly to the previous calculation for $^{11}$Be, $^{14}$Be and $^{14}$B \cite{1}.
The reaction corresponds to a neutron  
initially bound in $^{11}$Li  which is then excited into  
an unbound state of $^{10}$Li, assuming that another nucleon has been  
emitted and stripped by the target, thus not detected in coincidence with the core. 
One of the results of Ref. \cite{1}, as far as the reaction model is concerned, was to show that projectile fragmentation invariant mass spectra depend very weakly on the incident energy and on
the neutron initial binding energy. Related to this was  the investigation of the validity of the  
sudden approximation and the accuracy necessary in calculating the phase shifts.

 In Fig.  2 the symbols with error bars are the experimental points from \cite{2}. The thin full curve is the calculated total spectrum, sum of the contributions from the s and p components of the initial wave function, with the parameters of Table \ref{o} and the experimental spectroscopic factors from Ref. \cite{ers}. Using the theoretical spectroscopic factors from Refs. \cite{Thomp96}, \cite{Vinh96} gives very small differences in the shape of the total spectrum which however become unnoticeable after convolution with the experimental resolution function. The s, p and d final continuum states have  the parameters of Table \ref{q1}. Notice that these parameters are perfectly consistent with those extracted from the data and given by Table 4 of Ref. \cite{2}.  The thick solid  curve is the same calculation after convolution with the experimental resolution function. The dashed curve is the calculation without the d-resonance after convolution while the thin  dashed curve  is the bare calculation  without the d-resonance contribution. The agreement between  the thick solid curve  and the data   is quite good and it shows the importance of including the d-resonance to reproduce the tail of the experimental spectrum. The Coulomb breakup spectrum from the s component of the initial state was also calculated but since it gives
 a contribution of less than 5\% to the peak of the spectrum, it has been omitted from the figures.

The individual transitions, bound to unbound from each initial  
state component to each possible  unbound state as indicated, are  shown in Fig. 3. We give also in Fig. 4 the comparison for the s to s transition of our calculation with the calculation according to the sudden formula Eq.   (20) of \cite{bhe} which we discussed at length and compared to our model in Ref. \cite{1}.  Both calculations were done with the {\it exact} phase shift, obtained by solving numerically the Schr\"odinger equation with the potential given by Eqs.   (\ref{pot0},\ref{pot1}).  Contrary to what it was found in  Ref. \cite{1} for $^{13}$Be, we find here that in the case of the $^{10}$Li virtual state, the scattering length is large enough, i.e. the state is close enough to threshold, to justify the use of the sudden approximation. Furthermore we have checked that the effective range formula (cf. discussion after Eq.  (41) of Ref. \cite{1}) gives in this case a very good fit to the exact phase shift. In fact, using it in Eq.   (20) of \cite{bhe} gives a curve almost indistinguishable from the dashed line in Fig.  4. The parameters obtained from the fit are: a$_s$=-14.8 fm, r$_e$= 7.5 fm. Notice that the scattering length value obtained instead as a$_s$=-$\lim_{k \to 0}{\delta_0\over k}$ and given in Table \ref{q1} was -17.2 fm. This is the kind of sensitivity one can find using the two different prescriptions for a$_s$. On the other hand we have checked that the value of r$_e$ from the fit, is consistent with the behaviour of our   continuum state potential, Eqs.   (\ref{pot0},\ref{pot1}). Such potential becomes indeed negligible, and consistent with zero, for r$>$r$_e$= 7.5 fm,  as prescribed for the applicability of the effective range theory.

As already found in Ref. \cite{1} for $^{11}$Be and  $^{13}$Be, we can confirm here with the $^{10}$Li example that, a part for the s to s transition, the excitation of resonances with $l>0$ in the continuum of unbound nuclei in projectile fragmentation reactions, is a final state effect due to n-core interaction, rather than a  process in which a bound component of the initial wave function becomes suddenly unbound.
The results in Fig. 3 show clearly that the population of continuum resonance is dominated by the contributions from the s-initial state, while the transitions p to p or p to d  are not large enough to explain the experimental spectrum. This is particularly clear for the d-resonance whose presence is necessary to explain the experimental spectrum tail but whose strength could not be justified by a d to d transition which would have  a very small amplitude.
Therefore the strength of the continuum resonances of a daughter nucleus does not reflect directly the occupation of the bound states  of same angular momentum in the mother nucleus. This is different from the {\it common wisdom} on the breakup of two-neutron halo nuclei, as discussed for example in the recent Ref. \cite{nun}.

In the case of $^{11}$Li the two neutrons are in the same  state for each component of the initial wave function. Therefore since the p and d wave functions have less pronounced tails, the stripping probability of one of the two neutrons, as discussed in Sec.  3 of Ref. \cite{1}, will naturally diminish the absolute value of the peaks due to transitions from these states, with respect to  peaks due to transition from the bound s-component. This effect is not taken into account at the moment in our numerical implementations of the model.  On the other hand our absolute cross sections should  be multiplied by a factor two to take into account the fact that the experimental data do not distinguish the two neutrons in the continuum. The absolute value of our cross section can be read from the  dashed curve in Fig. 3. Taking into account the factor two just mentioned, we see that in order to compare to the experimental data in Fig. 2, which are given on their absolute scale, we still have to renormalise our calculations by a factor two. Considering the incertitude in the value of the neutron-target delta-interaction potential   and in the strong absorption radius, we can consider our estimates quite reasonable.
 
\section{Conclusions and outlook}

In this paper we have studied the energy spectrum of  $^{10}$Li as obtained from the fragmentation of $^{11}$Li by  applying a model \cite{1} for one neutron  
excitations from a bound initial state to an unbound resonant state in  
the
neutron-core low energy continuum. 
The model, based on a time dependent perturbation theory amplitude,  was  previously used \cite{1}  to  study  $^{13}$Be
and  the breakup of $^{11}$Be and it proved to be reliable. The same conclusion can be drawn here after comparing our calculations with the new data from \cite{2}.

 The initial state spectroscopic factors in  $^{11}$Li  are  quite well known experimentally, therefore the absolute values of our cross sections have also been checked, besides the shape of the n-core relative energy spectrum. Due to the closeness to threshold of the s virtual state we have seen that the sudden formula used in Ref. \cite{bhe} and the effective range approximation to the  phase shift, are both very   well justified for $^{10}$Li, contrary to what it was previously \cite{1} found for $^{13}$Be.
 Finally we have found that, in agreement with the interpretation given by the authors of \cite{2}, their recent data provide a clear evidence  on the excitation of a d resonance around 1.5 MeV. Such a resonance does not play much role in the composition of the $^{11}$Li ground state but it is an important building bloc of its excited states.

{\bf Acknowledgments}

\noindent We wish to thank Bj\"orn Jonson and his collaborators,  in particular Leonid Chulkov and Haik Simon, for communicating their results previous to publication and for  an enlightening correspondence.

\end{document}